%% file: mccs_oscillator_2010-03-11_v16-subm2-prb.tex
\documentclass[prb,twocolumn,eqsecnum,superscriptaddress,showpacs,letterpaper]{revtex4}
\usepackage{times}
\usepackage{verbatim}
\usepackage{bm}
\usepackage{bbm}
\usepackage{pifont}
\usepackage{amsmath}
\usepackage{amsfonts}
\usepackage{amscd}
\usepackage{amsthm}
\usepackage{amssymb}
\usepackage{hyperref}

\input{_macros-math-revtex.tex}
\input{_macros-common.tex}
\input{_macros-phys.tex}

\def\ket#1{\mathinner{|{#1}\rangle}}

\providecommand{\abs}[1]                {\lvert #1\rvert}

\renewcommand{\tfracs}[2]               {{#1}/{#2}}                          
\renewcommand{\tfracsb}[2]              {({#1}/{#2})}                        
\renewcommand{\tfraci}[2][]             {{#1}{#2}^{-1}}                      
\renewcommand{\tfddau}[2]               {{\partial #1}/{\partial #2}}        

\newcommand{\eqn}                       {Eq.~}
\newcommand{\eqns}                      {Eqs.~}
\newcommand{\sect}                      {Sec.~}
\renewcommand{\SU}                      {\ensuremath\mathrm{SU}}

\renewcommand{\vr}                      {\vec{r}}
\newcommand{\vq}                        {\vec{q}}
\newcommand{\ja}                        {{j_{\alpha}}}
\newcommand{\Na}                        {{N_{\alpha}}}
\newcommand{\phys}                      {\phi_\mathrm{phys}}

\newcommand{\AACS}[1]                   {\vec{A}_{#1}^{\CS}}           
\newcommand{\ACS}[1]                    {A_{#1}^{\CS}}                 
\newcommand{\dAACS}[1]                  {\delta \AACS{#1}}            
\newcommand{\dACS}[1]                   {\delta \ACS{#1}}             
\newcommand{\BBCS}[1]                   {\vec{B}_{#1}^{\CS}}           

\newcommand{\psiCS}[1]                  {\psi_{#1}^\CS}                     
\newcommand{\psiCSdagger}[1]            {\psi_{#1}^{\CS\,\dagger}}          

\newcommand{\AAstar}                    {\vec{A}^*}                          
\renewcommand{\ao}                      {a^\circ}

\newcommand{\aao}                       {\vec{a}^\circ}
\newcommand{\abar}                      {\wbar{a}}
\newcommand{\Po}                        {P^\circ}
\newcommand{\PPo}                       {\vec{P}^\circ}
\newcommand{\Pbar}                      {\wbar{P}}

\newcommand{\eperp}[1]                  {\vec{e}^\bot_{#1}}
\newcommand{\epar}[1]                   {\vec{e}^\parallel_{#1}}

\begin{document}

\title{Chern--Simons theory of multi-component quantum Hall systems}
\author{W. Beugeling}
\affiliation{Institute for Theoretical Physics, Utrecht University, Leuvenlaan 4, 3584 CE Utrecht, The Netherlands}
\author{M. O. Goerbig}
\affiliation{Laboratoire de Physique des Solides, CNRS UMR 8502, Universit\'e Paris Sud, F-91405 Orsay Cedex, France}
\author{C. Morais Smith}
\affiliation{Institute for Theoretical Physics, Utrecht University, Leuvenlaan 4, 3584 CE Utrecht, The Netherlands}
\date{\today}

\begin{abstract}
The Chern--Simons approach has been widely used to explain fractional quantum Hall states in the framework of trial wave functions. In the present paper, we generalise the concept of Chern--Simons transformations to systems with any number of components (spin or pseudospin degrees of freedom), extending earlier results for systems with one or two components. We treat the density fluctuations by adding auxiliary gauge fields and appropriate constraints. The Hamiltonian is quadratic in these fields and hence can be treated as a harmonic oscillator Hamiltonian, with a ground state that is connected to the Halperin wave functions through the plasma analogy. We investigate conditions on the coefficients of the Chern--Simons transformation and on the filling factors under which our model is valid. Furthermore, we discuss several singular cases, associated with states with ferromagnetic properties.
\end{abstract}

\maketitle

\begin{section}{Introduction}
In understanding the fractional quantum Hall effect (FQHE), the now famous trial wave function proposed by Laughlin\cite{Laughlin1983} proved to be a successful approach to describe the physics of incompressible quantum liquids at certain fractional filling factors.
Laughlin's wave function is furthermore the inevitable starting point for several generalisations, such as Jain's
composite-fermion proposal,\cite{Jain1989,Jain1990} Halperin's two-component wave function,\cite{Halperin1983} or more complicated
wave functions describing states possessing quasi-particle excitations with nonabelian statistics.\cite{MooreRead1991-ReadRezayi1999}

\par A field-theoretical approach, complementary to the above-mentioned one, consists of so-called Chern--Simons theories, which formalise the idea of flux attachment that is also implicit in the trial wave functions. Chern--Simons theories have been successfully elaborated to study incompressible\cite{LopezFradkin1991} and compressible\cite{HalperinEA1993} quantum liquids in one-component systems as well as two-component systems,\cite{WenZee1991PRB44,WenZee1992PRB46-WenZee1992PRL69,LopezFradkin1995,Rajaraman1997,ScarolaJain2001}
which comprise e.g.\ bilayer quantum Hall systems or single layer systems in situations where the spins are not completely polarised.
Multi-component Chern--Simons approaches have also been proposed in the study of edge excitations of the incompressible quantum Hall liquids.\cite{Orgad2007} An undeniable advantage of these Chern--Simons theories consists of their transparent insight into the exotic properties of these
quantum liquids, such as their topological degeneracy, the fractional charges of their quasi-particle excitations or the statistical properties of the latter.\cite{WenZee1991PRB44,WenZee1992PRB46-WenZee1992PRL69} However, the Chern--Simons theories are usually less adapted when it comes to calculating quantities involving energy scales. Indeed, Chern--Simons transformations act on the kinetic part of the electronic Hamiltonian, whereas they leave the interaction part invariant. The kinetic part gets therefore renormalised but continues to determine the overall energy scale whereas the physical energy scale in the FQHE must be set by the electron-electron interactions.

\par A successful generalisation of Chern--Simons theories, that does not suffer from the problem of the correct energy scale, is the Hamiltonian theory proposed by Shankar and Murthy.\cite{ShankarMurthy1997,MurthyShankar1998,Shankar1999,Shankar2001,MurthyShankar2003}
This theory is a very powerful tool for the computation of physical quantities, \cite{Shankar2001,MurthyShankar2003} and even for the description of higher-generation composite fermion states.\cite{GoerbigEA2004EPL-GoerbigEA2004PRB-GoerbigEA2006} However, it is limited by the fact that it does not incorporate internal degrees of freedom. The success of the single-component Hamiltonian theory justifies a generalisation that can be applied to describe systems for which internal degrees of freedom (spin and/or pseudospin) are relevant. The main interest in such a generalisation
stems from realistic systems with more than two internal degrees of freedom, such as graphene with its four-fold spin-valley degeneracy\cite{CastroNetoEA2009} or bilayer quantum Hall systems with non-polarised electron spins.

\par In this paper, we analyse a multi-component Chern--Simons theory within the framework of the microscopic theory by Shankar and Murthy.\cite{ShankarMurthy1997,MurthyShankar1998,Shankar1999,Shankar2001,MurthyShankar2003} This approach has two main advantages over the previously proposed ones. First,
it allows one to distinguish between physically relevant Chern--Simons theories from those which are ill-defined. The basic ingredient for this distinction is the $\kappa\times\kappa$ charge matrix $K$, which was first introduced by Wen and Zee.\cite{WenZee1991PRB44,WenZee1992PRB46-WenZee1992PRL69} We find that matrices with negative eigenvalues need to be discarded in the study of physically relevant Chern--Simons theories because they would lead to ground-state wave functions that cannot be normalised. This structural feature of Chern--Simons theories finds its physical interpretation within Laughlin's plasma
analogy\cite{Laughlin1983} that indicates a tendency of the different components to undergo a phase separation and thus to form spatially inhomogeneous states. We show that zero eigenvalues of the charge matrix $K$, in contrast to the unphysical negative eigenvalues, find a compelling interpretation in terms of ferromagnetic quantum Hall states.
Our results thus generalise previous work on two-component systems
by Lopez and Fradkin\cite{LopezFradkin1995} to an arbitrary number of components $\kappa$.

\par A second advantage of the present approach consists of a transparent
connection between multi-component Chern--Simons theories with trial wave
functions. It has been shown, in the simpler one-component case, that treating the fluctuations of the Chern--Simons vector
potential within the harmonic approximation (Gaussian model) yields Laughlin's and Jain's (unprojected) composite-fermion
wave functions.\cite{KaneEA1991,MurthyShankar2003} Similarly, we obtain here, within the Gaussian model of $\kappa$-component
fluctuating Chern--Simons vector potentials, multi-component trial wave functions\cite{GoerbigRegnault2007} that are generalisations
of Halperin's two-component wave functions.\cite{Halperin1983,LopezFradkin1995} Furthermore, we obtain in the same manner composite-fermion-type
wave functions that may be viewed as particular multi-component generalisations of Jain's original proposal.\cite{Jain1989,Jain1990}

\par The paper is organised as follows. In \sect\ref{sect_cstransformations}, we define the Chern--Simons transformations for systems with $\kappa$ components and introduce, in \sect\ref{sect_hamiltonian_theory}, extra degrees of freedom, in the form of the auxiliary gauge fields, as described by Shankar and Murthy. We subsequently diagonalise the harmonic oscillator Hamiltonian and investigate the connection of the resulting wave function with trial wave functions through the plasma analogy. In \sect\ref{sect_singular}, we extend our results to the situation of singular $K$ matrices and discuss the relation between residual
symmetries and underlying ferromagnetic properties of the quantum Hall states.
Our conclusions are presented in \sect\ref{sect_conclusions}.
\end{section}

\begin{section}{Chern--Simons transformations}
\label{sect_cstransformations}
We consider a quantum Hall system with $\kappa$ internal states, hereafter referred to as ``components''. In the simplest case of a two-dimensional electron gas at a GaAs/AlGaAs interface, one has $\kappa=2$ for the two possible orientations of the electron spin. The case $\kappa=4$ is relevant for bilayer quantum Hall systems, where a second pseudospin mimics the layer index, or in graphene due to its two-fold valley degeneracy, in addition to the physical spin of the electrons. Higher values of $\kappa$ are rarely discussed in the literature, but may play a role in the context of multilayer systems or of bilayer graphene, where the
zero-energy level consists of the $n=0$ and $n=1$ Landau levels.\cite{CastroNetoEA2009}
The Chern--Simons transformation\cite{ArovasEA1985,Jain1989} is defined by the relation between the $\kappa$ original electronic fields $\psi_{\alpha}(\vr)$ and the $\kappa$ transformed fields $\psiCS{\alpha}(\vr)$ as
\begin{align}\label{eqn_cs_definition}
  \psi_\alpha(\vr)
  &=\exp\biggl(
    -\ii\int\intd^2\vr'\theta(\vr-\vr')\sum_{\beta=1}^\kappa
    K_{\alpha\beta}\rho_\beta(\vr')
  \biggr)\psiCS{\alpha}(\vr),
\end{align}
where $\theta(\vr) = \arg(x+\ii y)$ indicates the angle between the vector $\vr=(x,y)$ and the $\unitvec{x}$
direction, and $\rho_\beta(\vr) = \psi^\dagger_\beta(\vr)\psi_\beta(\vr)=\psiCSdagger{\beta}(\vr)\psiCS{\beta}(\vr)$ is the density operator of the particles of
component $\beta$. The $\kappa\times \kappa$ matrix $K_{\alpha\beta}$ encodes the topological
properties of the underlying quantum liquids, such as its degeneracy, the charges of its quasi-particle excitations and the
statistics of the latter.\cite{WenZee1991PRB44,WenZee1992PRB46-WenZee1992PRL69} Physically, it indicates the number of flux quanta
attached to particles of component $\alpha$ due to the density of particles of component $\beta$. This transformation is a \emph{singular} transformation for the reason that $\theta(\vr-\vr')$ has a singularity at $\vr'=\vr$.

\par The gauge transformation is defined such that it generates the gauge potentials\footnote{Throughout the text, all sums over the
component indices ($\alpha, \beta, \ldots$) are over the values $1,\ldots,\kappa$, unless indicated otherwise.}
\begin{align}\label{eqn_cs_vectorpotential}
  \AACS{\alpha}(\vr)=-\frac{\hbar}{e}\vec{\nabla}_\vr\int\intd^2\vr'\theta(\vr-\vr')\sum_\beta
  K_{\alpha\beta}\,\rho_\beta(\vr'),
\end{align}
and such that the one-particle Hamiltonian $[-\ii\hbar\nabla+e\,\vec{A}(\vr)]^2/2m$ for the component $\alpha$
is transformed to
\[
  H_{\alpha} = \frac{1}{2m}\left[-\ii\hbar\nabla+e\,\vec{A}(\vr) + e\,\AACS{\alpha}(\vr)\right]^2.
\]
Here, $m$ is the mass of the particles, and $e$ is the electron charge.
By using $\nabla\times\nabla\theta(\vr) = 2\pi\,\delta(\vr)$, we derive the corresponding magnetic fields,
\[
  \BBCS{\alpha}(\vr) = -\frac{h}{e}\sum_\beta K_{\alpha\beta}\,\rho_\beta(\vr)\unitvec{z}.
\]

\par Since $\AACS{\alpha}$ is a gauge field, its Fourier transform $\AACS{\alpha}(\vq)$ may be fixed to a convenient gauge. We choose it to be transverse, $\ii\vq\cdot\AACS{\alpha}(\vq)=0$, so that it fixes the direction of $\AACS{\alpha}(\vq)$ to be $\unitvec{z}\times\vq/\abs{\vq}$, up to a sign. For the magnitude, we use that under a Fourier transform $\vec{B}(\vr)=\nabla \times \vec{A}(\vr)$ transforms to $\vec{B}(\vq)=\ii\vq\times \vec{A}(\vq)$, such that we obtain
\begin{equation}\label{eqn_vector_potential_fourier}
  \AACS{\alpha}(\vq)
  =\ACS{\alpha}(\vq)\eperp{\vq}
  =-\frac{h}{e\abs{\vq}}\sum_\beta K_{\alpha\beta}\,\rho_\beta(\vq)\eperp{\vq},
\end{equation}
where we define the transverse unit vector as $\eperp{\vq}=\ii\unitvec{z}\times \vq/\abs{\vq}$.

\par The effective magnetic field seen by the composite particles of type $\alpha$ is
\begin{align}\label{eqn_bstar}
  \vec{B}^*_\alpha
   &= \vec{B} + \avg{\BBCS{\alpha}}
   = B\Bigl(1-\sum_\beta K_{\alpha\beta}\nu_{\beta}\Bigr)\unitvec{z},
\end{align}
where $\nu_\beta$ are the component filling factors, given by $\nu_\beta =
\tfracsb{h}{eB}\nel{\beta}=2\pi\lB^2\nel{\beta}$ in terms of the electronic densities $\nel{\beta}$ and of the magnetic length $\lB=\sqrt{\hbar/eB}$. This result is an extension of the two-component case
presented in Refs.~\onlinecite{Rajaraman1997,ScarolaJain2001}.
Notice that each particle type has its own effective magnetic field, and
hence also its own magnetic length $\lBB{B^*_\alpha} = \sqrt{\hbar/eB^*_\alpha}$. The composite particle
filling factors $\nu^*_\alpha$
are expressed in terms of the electronic filling factors $\nu_\alpha$ as\cite{Rajaraman1997}
\begin{align}\label{eqn_lbstar}
  \frac{\nu^*_\alpha}{\nu_\alpha}
  &=
    \frac{\lBB{B^*_\alpha}^2}{\lB^2}
    = \frac{B}{B^*_\alpha}
    = \frac{1}{1-\sum_\beta K_{\alpha\beta}\,\nu_{\beta}}.
\end{align}
This result generalises the one-component relation
\begin{equation}\label{eqn_nustar_onecomponent}
  \nu^*=\frac{\nu}{1-2s\nu} \qquad \leftrightarrow \qquad \nu=\frac{\nu^*}{2s\nu^*+1},
\end{equation}
in terms of the Chern--Simons charge $K=2s$.

\par The statistical angle associated with the exchange of the transformed fields $\psiCS{\alpha}$ and $\psiCSdagger{\alpha}$ can be derived by using their definition, \eqn\eqref{eqn_cs_definition}, and the fact that the original fields are fermionic.
Under the condition that the charge matrix $K_{\alpha\beta}$ is symmetric, which is a generalisation of the condition discussed in the two-component
case,\cite{Rajaraman1997} we obtain
\[
  \psiCS{\alpha}(\vr_1)\psiCS{\beta}(\vr_2)
   + \ee^{\ii\pi\,K_{\alpha\beta}}\psiCS{\beta}(\vr_2)\psiCS{\alpha}(\vr_1)
  = 0
\]
and
\begin{multline*}
  \psiCS{\alpha}(\vr_1)\psiCSdagger{\beta}(\vr_2)
   + \ee^{\ii\pi\,K_{\alpha\beta}}\psiCSdagger{\beta}(\vr_2)\psiCS{\alpha}(\vr_1)\\
  = \delta_{\alpha\beta}\,\delta(\vr_1-\vr_2).
\end{multline*}
Thus, we have found that the statistical angles of the exchange are $\pi K_{\alpha\beta}$, i.e., proportional to the entries of the charge matrix.
The parity of the diagonal elements $K_{\alpha\alpha}$ of the charge matrix $K$ determines the statistical
properties of the Chern--Simons fields $\psiCS{\alpha}$. If they are even integers, the originally fermionic electron fields
$\psi_{\alpha}$ are transformed into fermionic Chern--Simons fields. However, one may also change the statistical properties
of the fields from fermions to bosons by using odd integers for the diagonal components $K_{\alpha\alpha}$.
In the following sections, we mainly discuss fermionic Chern--Simons fields, in order to make a connection with the composite-fermion
theory, although the main conclusions of the paper also apply to bosonic fields.
\end{section}

\begin{section}{Gaussian theory}
\label{sect_hamiltonian_theory}
\begin{subsection}{Auxiliary gauge fields}
\label{sect_aux_gauge_fields}%
The formalism proposed by Shankar and Murthy\cite{MurthyShankar1998, MurthyShankar2003} allows us to
treat the fluctuations of the Chern--Simons vector potential
via the introduction of $\kappa$ real-valued transverse gauge fields $\aao_{\alpha}(\vr_\ja)$.\cite{BohmPines1953}
The extended Chern--Simons Hamiltonian in first quantisation, with $N_\alpha$ particles of each type
$\alpha$, reads
\begin{multline}\label{eqn_hamiltonian_firstquant}
  H_{CS}
  =\frac{1}{2m}\sum_\alpha\sum_{\ja=1}^{\Na}
    \bigl[\vec p_\ja + e\,\AAstar_{\alpha}(\vr_\ja)\bigr.\\
    \bigl.+ e\,\delta\AACS{\alpha}(\vr_\ja) + e\,\aao_{\alpha}(\vr_\ja)\bigr]^2,
\end{multline}
where we absorb the average value of the Chern--Simons potential \eqref{eqn_cs_vectorpotential} into an effective vector potential
$\AAstar_\alpha(\vr) = \vec A(\vr)+\avg{\AACS{\alpha}}$. This definition
yields the effective magnetic field
$\nabla\times\AAstar_\alpha(\vr_\ja)=\vec{B}^*_\alpha(\vr_\ja)$ given in
\eqn\eqref{eqn_bstar}. In Fourier space, the fluctuations $\dAACS\alpha(\vq)$ are transverse, similar to the gauge field itself, as given by \eqn\eqref{eqn_vector_potential_fourier}. Here, we have
\begin{align*}
  \dAACS\alpha(\vq)= \dACS\alpha(\vq)\eperp{\vq}
   = \frac{h}{e\abs{\vq}}\sum_\beta K_{\alpha\beta}\,\delta\rho_\beta(\vq)\eperp{\vq}.
\end{align*}

\par Since we have artificially added the auxiliary gauge field $\aao_{\alpha}(\vr)$, we have enlarged the
Hilbert space, where the physical states form only a subspace $\{\ket{\phys}\}$
characterised by
\begin{equation}\label{eqn_constraint_cs}
  \ao_\alpha(\vq)\ket{\phys} = 0,
\end{equation}
for all components $\alpha$. In other words, the gauge field operator acting on any physical state vanishes.

\par Additionally, we introduce a longitudinal field $\PPo(\vq)=\ii\Po(\vq)\epar{\vq}$ (with $\epar{\vq}\equiv\vq/\abs{\vq}$), conjugate and perpendicular to the newly introduced gauge field  $\aao(\vq)=\ao(\vq)\eperp{\vq}$, according to the commutation
relation in Fourier space
\[
  [\ao_\alpha(\vq),\Po_\beta(-\vq')]=\ii\hbar\delta_{\alpha\beta}\delta_{\vq,\vq'}.
\]
Since the operator $\Po_\alpha$ is conjugate to $\ao_\alpha$, it
generates translations in $\ao_\alpha$,
as may be seen from the definition
\[
  U
  =\exp\biggl(\frac{\ii}{\hbar}\sum_\alpha\sum_{\vq'}\Po_\alpha(-\vq')\dACS{\alpha}(\vq')\biggr),
\]
which translates $\ao_\beta$ by the vector $-\dACS{\beta}(\vq)$ as
$U^\dagger\ao_\beta(\vq)U = \ao_\beta(\vq) - \dACS{\beta}(\vq)$.
By using this shifting property of $U$, which is also valid in $\vr$-space,
and with
$[\vec{p}_\ja, U]=\tfracsb{h}{eL^2}\sum_\beta K_{\alpha\beta}\PPo_\beta(\vr_\ja)$,
we may eliminate $\dAACS{\beta}(\vq)$ from the Hamiltonian of \eqn\eqref{eqn_hamiltonian_firstquant}, which then transforms into
\begin{align*}
  H_{\CP}
  &=U^\dagger H_\CS U
  =\frac{1}{2m}\sum_\alpha\sum_{\ja=1}^{\Na}
    \biggl[
      \vec p_\ja + e\,\AAstar_{\alpha}(\vr_\ja) \biggr.\\
  &\hspace{60.2pt} \biggl.{} + e\,\aao_{\alpha}(\vr_\ja)+ \frac{h}{eL^2}\sum_\beta K_{\alpha\beta}\PPo_\beta(\vr_\ja)
    \biggr]^2,
\end{align*}
while transforming the states to $\psi^\CP = U\inv \psi^\CS$. In these equations, $L^2$ is the area of the
system. By transforming the states, we also transform the
constraint \eqref{eqn_constraint_cs} to
\begin{align}\label{eqn_constraint_cp}
  \lefteqn{\bigl(\ao_\alpha(\vq)-\dACS{\alpha}(\vq)\bigr)\ket\phys}&{}\nonumber\\
  &\quad{}= \biggl(\ao_\alpha(\vq) - \frac{2\pi\hbar}{e\abs{\vq}}\sum_\beta K_{\alpha\beta}\,\delta\rho_\beta(\vq) \biggr)\ket\phys
  = 0.
\end{align}

\par The Hamiltonian may be decomposed into three terms, $H_\CP=H^*+H_{\mathrm{coupl}}+H_{\mathrm{aux}}$, given by
\begin{align}\label{eqn_hstar}
  H^* &=
  \frac{1}{2m}\sum_\alpha\sum_{\ja=1}^{\Na}\vec{\Pi}_\ja^2,\\
\label{eqn_hcoupl}
  H_{\mathrm{coupl}}
  &=\frac{1}{m}\sum_\alpha\sum_{\ja=1}^{\Na}\!\vec\Pi_\ja\!\cdot\!
  \biggl[e\aao_{\alpha}(\vr_{\ja})
  \!+\!b\sum_\beta K_{\alpha\beta}\PPo_\beta(\vr_\ja)\biggr],\\
\label{eqn_hosc_exact}
 H_{\mathrm{aux}}
  &= \frac{1}{2m}\sum_\alpha\sum_{\ja=1}^{\Na}\biggl[
  e^2 \ao_{\alpha}{}^2(\vr_{\ja})\biggr.\\
  &\hspace{50.1pt}{}+\biggl.b^2 \sum_\beta\sum_\gamma\Po_{\beta}(\vr_{\ja})K_{\beta\alpha} K_{\alpha\gamma}\Po_{\gamma}(\vr_{\ja})
  \biggr]\nonumber,
\end{align}
where $\vec\Pi_\ja\equiv\vec p_{\ja}+e\vec A_{\alpha}^*(\vr_{\ja})$ and $b=\tfracs{h}{eL^2}$, which has the dimensions of a magnetic field. Notice that for $H_{\mathrm{aux}}$ we have used that $\aao_\alpha(\vq)$ and $\PPo_\beta(\vq)$ are perpendicular.

\par In the remainder of this paper, we discuss only the term $H_\mathrm{aux}$ that involves the auxiliary gauge fields. The full theory, including the other terms of the Hamiltonian shall be discussed in a future publication.\cite{BeugelingEA2009IP2(hamtheory)}
\end{subsection}

\begin{subsection}{Gaussian model of the auxiliary gauge fields}\label{sect_harmonic_oscillator}
We shall now analyse $H_{\mathrm{aux}}$ in detail. By observing that $\sum_\ja\delta(\vr-\vr_\ja) =
\rho_\alpha(\vr)=\nel{\alpha}+\delta\rho_\alpha(\vr)$, we can rewrite \eqn\eqref{eqn_hosc_exact} as
\begin{multline*}
 H_{\mathrm{aux}}
  = \frac{1}{2m}\sum_\alpha\int\intd^2\vr\,\rho_\alpha(\vr)\,\biggl(
   e^2 \ao_{\alpha}{}^2(\vr)\biggr.\\
  \biggl.{}+b^2 \sum_\beta\sum_\gamma\Po_{\beta}(\vr)K_{\beta\alpha}K_{\alpha\gamma}\Po_{\gamma}(\vr)
  \biggr).
\end{multline*}
Up to this point, all equations are exact. Now, we approximate $H_\mathrm{aux}$ by assuming that the density
fluctuations $\delta\rho_\alpha$ are small with respect to the average densities $\nel{\alpha}$. Since the resulting Hamiltonian
becomes quadratic, this approximation is called the
\emph{harmonic approximation}. We should keep in mind that this approximation breaks down if the fluctuations are not
small with respect to the average densities. In particular, the approximation is certainly invalid if one of the
average densities is zero. We therefore assume that none of the average densities $\nel{\alpha}$
vanishes. However, in the case of a singular charge matrix $K$, a redefinition of the filling factors might lift this problem, as will be discussed in more detail in \sect\ref{sect_singular}. The Hamiltonian $H_{\mathrm{aux}}$ in Fourier space is approximated by
\begin{multline}\label{eqn_hosc}
  H_{\mathrm{osc}}
  =\sum_{\vq}\sum_\alpha\frac{\nel{\alpha}L^2}{2m}
  \biggl(
   e^2\ao_{\alpha}(-\vq)\ao_{\alpha}(\vq)\biggr.\\
  \biggl.{}+b^2 \sum_\beta\sum_\gamma\Po_{\beta}(-\vq)K_{\beta\alpha}K_{\alpha\gamma}\Po_{\gamma}(\vq)
  \biggr),
\end{multline}
where we note that $\ao(-\vq)=(\ao(\vq))^\dagger$ and $\Po(-\vq)=(\Po(\vq))^\dagger$. Because the Hamiltonian \eqref{eqn_hosc} is
quadratic in the gauge fields $\ao_\alpha$ and its conjugate fields $\Po_\alpha$, it is possible to write it in terms of ladder operators. However, due to the appearance of the
matrices $K$ in the term with $\Po$'s, it is a nontrivial task to define suitable ladder
operators $\A_\alpha(\vq)$ such that the commutators between them are of the form
$\comm{\A_\alpha(\vq)}{\A^\dagger_\beta(\vq')} = \delta_{\alpha\beta}\delta_{\vq,\vq'}$.

\par In order to diagonalise the Hamiltonian, we define $N=\diag(\{\nu_{\alpha}\})$ as the dimensionless diagonal matrix of filling factors,
$N_{\alpha\beta}=\nu_\alpha\delta_{\alpha\beta}$, and also write the fields and their conjugates as vectors in the component space,
$\ao=(\ao_1,\ldots,\ao_\kappa)$ and $\Po=(\Po_1,\ldots,\Po_\kappa)$. We omit the $\vq$ dependence for a while. In this concise notation, the oscillator Hamiltonian can be written as
\begin{equation}\label{eqn_hosc_matform}
  H_\mathrm{osc}
  = \frac{L^2}{2m}\frac{eB}{h}\left[e^2\ao{}^\dagger N\,\ao
   + b^2\Po{}^\dagger\,K^\dagger NK\,\Po\right].
\end{equation}
The prefactor can also be written as $\tfracs{L^2\omegac}{2h}$, where $\omegac=\tfracs{eB}{m}$ is the cyclotron
frequency. We recall that the matrix $K$ is real and symmetric, so that $K^\dagger=K$. We perform the
diagonalisation in two steps. First, we define $a'=\sqrt{N}\ao$ and $P' = \sqrt{N\inv}\Po$, so that the Hamiltonian becomes
\begin{equation*}
  H_\mathrm{osc}
  = \frac{L^2\omegac}{2h}\left[e^2 a'{}^\dagger a'
   + b^2 P'{}^\dagger\sqrt{N}\,KNK\sqrt{N}P'\right].
\end{equation*}
The matrix between the $P'$'s is the square of the matrix $E\equiv\sqrt{N}\,K\sqrt{N}$, which is
real and symmetric. Therefore, it can be diagonalised in terms of a diagonal matrix $D$ and an orthogonal matrix $C$, such that $E = C\inv D\,C$. The matrix $D$ has the eigenvalues $\lambda_\alpha$ of $E$ on its diagonal, and $C\Tp$ contains the corresponding eigenvectors as columns. The ability to choose $C$ as an orthogonal matrix (i.e., $C\inv=C\Tp$) is provided by the property that the matrix $E$ is symmetric, so that the eigenvectors can be chosen such that they form an orthonormal basis. Having found the diagonalisation $E = C\Tp D\,C$, we define
\begin{equation}\label{eqn_abar_pbar}
  \abar=C\,a'=C\sqrt{N}\ao,
  \quad
  \Pbar= C\, P'=C\sqrt{N\inv}\Po,
\end{equation}
so that the Hamiltonian becomes
\begin{subequations}\label{eqn_hosc_diagonal_}
\begin{align}
  H_\mathrm{osc}
  &=\frac{L^2\omegac}{2h}\left[ e^2\abar^\dagger\abar +b^2 \Pbar^\dagger D^2\,\Pbar\right]
  \label{eqn_hosc_diagonal__matrixform}\\
  &=\frac{L^2\omegac}{2h}\sum_\alpha\biggl[
  e^2\abar^\dagger_\alpha\abar_\alpha + b^2 \Pbar^\dagger_\alpha\lambda_\alpha^2\,\Pbar_\alpha
  \biggr],
  \label{eqn_hosc_diagonal__components}
\end{align}
\end{subequations}
written in matrix form and in components, respectively. For the derivation we have used that $\sum_\alpha\abar^\dagger_\alpha\abar_\alpha=\abar^\dagger\abar=a'{}^\dagger a'$ by virtue of the orthogonality of $C$, $\sum_\gamma C_{\alpha\gamma}C_{\beta\gamma}=\sum_\gamma C_{\alpha\gamma}C\Tp_{\gamma\beta}=\delta_{\alpha\beta}$.
For this transformation to be well-defined, it is required that $\nu_{\beta}\not=0$ for all components
$\beta$, which we already assumed in order for the harmonic approximation to be valid.
The definition is such that the
commutator between $\abar$ and $\Pbar$ is given by
\begin{align}
  \comm{\abar_\alpha(\vq)}{\Pbar_\beta(-\vq')}
  =\ii\hbar\,\delta_{\alpha\beta}\delta_{\vq,\vq'},
  \label{eqn_comm_abar_pbar}
\end{align}
which holds also by virtue of the orthogonality of $C$.

\par By setting $\Po_\alpha = -\ii\hbar\ddau{}{\ao_\alpha}$ and consequently $\Pbar_\alpha = -\ii\hbar\ddau{}{\abar_\alpha}$, we
can derive that
\begin{equation}\label{eqn_chi_osc1}
  \chi_\mathrm{osc}=\exp\biggl(-\frac{e}{2\hbar b}\sum_{\vq}\sum_\alpha \abar_{\alpha}(-\vq)\xi_\alpha\abar_{\alpha}(\vq)\biggr)
\end{equation}
is a ground state of the Hamiltonian \eqref{eqn_hosc_diagonal_} if we set $\xi_\alpha = \tfraci{\abs{\lambda_\alpha}}$. Evidently, $\xi_\alpha$ is only well defined if the matrix $E$ is nonsingular, i.e., if none of its eigenvalues is zero. Moreover, the eigenvalues appearing in the eigenstate are actually not the eigenvalues of $E$ itself, but the square
roots of the eigenvalues of $E^2=\sqrt{N}\,KNK\sqrt{N}$, namely $\sqrt{\lambda_\alpha^2} = \abs{\lambda_\alpha}$.
The ground state \eqref{eqn_chi_osc1} can then be written in matrix form as
\begin{align}
  \chi_\mathrm{osc}
  &=\exp\biggl(-\frac{e}{2\hbar b} \abar^\dagger D\inv \abar\biggr)
  =\exp\biggl(-\frac{e}{2\hbar b}\ao{}^\dagger K^{-1}\ao\biggr)
  \label{eqn_chi_osc1_2_matrixform}
\end{align}
where we used $\abar^\dagger D\inv \abar=\ao{}^\dagger K\inv\ao$ in order to write the ground state
in terms of the original auxiliary gauge fields $\ao$.
Notice that, had we chosen the negative square roots $-\sqrt{\lambda_\alpha^2}$ for the eigenvalues of $E$, the ground-state
wave function \eqref{eqn_chi_osc1_2_matrixform} could not be normalised. Negative eigenvalues are indeed unphysical because they
would lead to an instability of the electron liquid, the components of which phase-separate, as may be seen within the plasma picture
of the FQHE.\cite{DeGailEA2008} It is therefore important, for the structure of the Chern--Simons theory to be well-defined, to discard
negative eigenvalues $\lambda_{\alpha}$. This is namely the case for the analysis presented in \sect\ref{sect_halperin_connection}, where we assume a positive definite $K$.
The case of zero eigenvalues is treated separately in \sect\ref{sect_singular}.

\par Acting with the Hamiltonian \eqref{eqn_hosc_diagonal_} on the ground state \eqref{eqn_chi_osc1} gives its energy
eigenvalues
\[
  \sum_\alpha\frac{L^2\omegac}{2h}\hbar e b\abs{\lambda_\alpha}
  = \frac{\hbar\omegac}{2}\sum_\alpha\abs{\lambda_\alpha}
  = \frac{\hbar}{2}\sum_\alpha\omega_\alpha,
\]
where $\omega_\alpha=\abs{\lambda_\alpha}\omegac$ are the characteristic frequencies, given in terms of the eigenvalues $\lambda_\alpha$ and the cyclotron frequency $\omegac$.

\par At this point, we define the ladder operators as
\begin{equation}\label{eqn_ladder_operators}
 \begin{aligned}\A_\alpha(\vq)
  &=\frac{L}{\sqrt{4\pi\hbar^2\lambda_\alpha}}\left(e\,\abar_\alpha(\vq) + \ii b\lambda_\alpha\Pbar_\alpha(\vq)\right),\\
  \A_\alpha^\dagger(\vq)
  &=\frac{L}{\sqrt{4\pi\hbar^2\lambda_\alpha}}\left(e\,\abar_\alpha(-\vq) -
  \ii b\lambda_\alpha\Pbar_\alpha(-\vq)\right),
  \end{aligned}
\end{equation}
still under the assumption that the eigenvalues $\lambda_\alpha$ are positive. The commutator of the rescaled ladder operators becomes
$\comm{\A_\alpha(\vq)}{\A^\dagger_\beta(\vq')}=\delta_{\alpha\beta}\delta_{\vq,\vq'}$, so that
$\A^\dagger_\alpha(\vq)\A_\alpha(\vq)$ is the number operator for the oscillator states in the component $\alpha$
of the diagonalised basis. The
Hamiltonian can be conveniently written in terms of the ladder operators as
\begin{align}
  H_\mathrm{osc}
  &=\sum_\vq\sum_\alpha
  \hbar\omega_\alpha\bigl(\A^\dagger_\alpha(\vq)\A_\alpha(\vq)+\tfrac{1}{2}\bigr).
\label{eqn_hosc_ladder}
\end{align}
This result also proves that the ``ground state'' \eqref{eqn_chi_osc1} is indeed the lowest-energy state.

\par Notice that the energies $\hbar \omega_{\alpha}$ play the role of \emph{quasi-particle gaps} in the Chern--Simons theory, and
the ground state is well-defined for $\det(K)\neq 0$.\cite{WenZee1991PRB44} Zero-energy gaps are obtained if one of the
eigenvalues $\lambda_{\alpha}=0$, i.e., when the matrix $K$ is singular, $\det(K)=\det(E)=0$. Contrary to what one may naively expect,
this situation is not in contradiction with an incompressible quantum liquid, where all (collective) charge modes must be
gapped. As we discuss in more detail in \sect\ref{sect_singular}, the zero-gap modes associated with $\lambda_{\alpha}=0$
reveal ferromagnetic properties of the underlying state,\cite{GoerbigRegnault2007} which in the presence of interactions evolve into spin-wave modes while keeping the charge modes gapped.
\end{subsection}

\begin{subsection}{Connection with trial wave functions}
\label{sect_halperin_connection}
In order to obtain the wave functions corresponding to the ground state \eqref{eqn_chi_osc1_2_matrixform},
we may rewrite it in terms of the density fluctuations $\delta\rho_\alpha(\vq)$,
using the constraint \eqref{eqn_constraint_cp}. Once again, it is more
convenient to do the computation in matrix notation. The constraint is then given by
$\ao=\tfracsb{h}{e\abs{\vq}}K\,(\delta\rho)$ for physical states, with $(\delta\rho) =
(\delta\rho_1,\ldots,\delta\rho_\kappa)$ the vector of the density fluctuations. Hence, we find
\begin{align}
  \chi_\mathrm{osc}
   &= \exp\biggl(-\frac{1}{2}(\delta\rho)^\dagger \frac{2\pi L^2}{\abs{\vq}^2}K\,(\delta\rho)\biggr).
   \label{eqn_chi_osc3_matrixform}
\end{align}
Notice that, written in terms of density fluctuations, the ground-state wave function is no longer confronted
with the problem of zero-eigenvalues of $E$ (or $K$) because it is the matrix $K$, and not its inverse $K^{-1}$, which appears here.

As shown in Ref.~\onlinecite{KaneEA1991}, we may relate the expression \eqref{eqn_chi_osc3_matrixform}
to the plasma picture proposed by Laughlin in his original publication.\cite{Laughlin1983}
In this picture, we regard $\abs{\chi_\mathrm{osc}}^2$ as the
Boltzmann weight $\exp(-\beta\mathcal{H})$ of the plasma Hamiltonian $\mathcal{H}$, where one sets $\beta=2$ (Ref.~\onlinecite{DeGailEA2008}). Then $\mathcal{H}$ can be identified as the Hamiltonian of particles interacting due to the Coulomb potential in two dimensions, $-\!\log\abs{\vr}$, which equals $2\pi L^2/\abs{\vq}^2$ in momentum space. As discussed in Appendix~\ref{sect_plasma_analogy_derivation}, the wave function that we obtain is
\begin{widetext}
\begin{equation}\label{eqn_wf_halperin1}
    \psi(\{z_\ja\})=
    \prod_{\alpha}\prod_{\substack{j_\alpha,k_\alpha\\j_\alpha < k_\alpha}}(z_{j_\alpha}-z_{k_\alpha})^{K_{\alpha\alpha}}
    \prod_{\substack{\alpha,\beta\\ \alpha<\beta}}\prod_{j_\alpha, k_\beta}(z_{j_\alpha}-z_{k_\beta})^{K_{\alpha\beta}}
    \exp\biggl(-\sum_{\alpha,\beta}\nu_{\alpha}K_{\alpha\beta}\sum_{k_\beta}\frac{\abs{z_{k_\beta}}^2}{4\lB^2}\biggr)
    \,\phi_{\{\nu_\alpha^*\}}(\{z_\ja\}),
\end{equation}
\end{widetext}
where we write $z=x-\ii y$. This wave function is a product of the oscillator function and the wave function
$\phi_{\{\nu_\alpha^*\}}(\{z_\ja\})$, which encodes the residual degrees of freedom for particles in the reduced field
${\bf B}_{\alpha}$, i.e., at the effective filling factors $\nu_{\alpha}^*$ given by \eqn\eqref{eqn_lbstar}. Quite generally,
one may describe the same system in the framework of different Chern--Simons theories, according to how much flux is absorbed
in the transformation by the matrix $K_{\alpha\beta}$. It is often convenient, if possible, to choose the Chern--Simons transformation
such that the residual wave function is factorisable into single-component wave functions $\tilde{\phi}_{\nu_\alpha^*}$,
\begin{equation}\label{eqn_factoris}
\phi_{\{\nu_\alpha^*\}}(\{z_\ja\})=\prod_{\alpha=1}^{\kappa}\tilde{\phi}_{\nu_\alpha^*}(\{z_\ja\}),
\end{equation}
so that each component may be treated independently after the transformation. Notice, however, that this aim may be in conflict
with the above-mentioned condition of positive eigenvalues of the charge matrix $K_{\alpha\beta}$, namely in the context of
symmetric states with ferromagnetic properties that we discuss in \sect\ref{sect_singular_harmosc}.

\par The simplest state of a factorisable residual wave function according to \eqn\eqref{eqn_factoris} consists of a product of states at an effective filling factor $\nu_{\alpha}^*=1$ for each component, each of which involves a Slater determinant, in the form
\begin{equation}\label{eqn_res_nu1}
\tilde{\phi}_{\nu_\alpha^*=1}(\{z_\ja\})=\prod_{j_\alpha < k_\alpha}(z_{j_\alpha}-z_{k_\alpha})
\exp\biggl(-\sum_{k_\alpha}\frac{\abs{z_{k_\alpha}}^2}{4\lBB{B^*_\alpha}^2}\biggr).
\end{equation}
Such a state would then correspond to a Halperin wave function that is described by an exponent matrix  $M_{\alpha\beta}=K_{\alpha\beta}+\delta_{\alpha\beta}$.
In order to
have a fermionic wave function, the elements $K_{\alpha\alpha}$ must naturally be even integers, and we thus have a Chern--Simons theory that transforms fermions into (composite) fermions. Alternatively, one may have chosen the bosonic version of the Chern--Simons theory, in which case the diagonal elements of the matrix $K_{\alpha\beta}=M_{\alpha\beta}$ would be odd. The same state \eqref{eqn_wf_halperin1} would then be described as a product of the oscillator wave function $\chi_\mathrm{osc}$, which absorbs all the flux, and a \emph{bosonic} wave function for zero net magnetic field $B_{\alpha}^*=0$, for all components, $\phi_{\{B_\alpha^*=0\}}(\{z_\ja\})=1$.

\par Until now, we have discussed states that may be described in terms of generalised $\kappa$-component Halperin wave functions,\cite{GoerbigRegnault2007} where the residual wave function $\phi_{\{\nu_\alpha^*\}}(\{z_\ja\})$ is itself a (typically simpler) Halperin wave function described by a ``residual'' exponent matrix $M_{\alpha\beta}^*$ such that $M_{\alpha\beta}=K_{\alpha\beta}+M_{\alpha\beta}^*$ (see also Appendix~\ref{sect_plasma_analogy_derivation}). Notice, however, that the Chern--Simons theory discussed above may also provide us with another class of factorisable trial wave functions
if we replace the Slater determinants \eqref{eqn_res_nu1} for the effective filling factors $\nu_{\alpha}^*=1$ by Slater determinants for $p_{\alpha}$ completely filled composite-fermion levels $\phi^{(\alpha)}_{p_\alpha}(\{z_\ja\})$ in each component. The resulting wave function \eqref{eqn_wf_halperin1} is related to the $\kappa$-component Halperin wave function in the same manner as Jain's one-component composite-fermion\cite{Jain1989,Jain1990} to Laughlin's wave function.\cite{Laughlin1983} Naturally, the proposed Slater determinants contain non-analytic components in the polynomial, and, in the same manner as for Jain's wave functions, one needs to project the resulting
wave function to the subspace of analytic functions in order to satisfy the lowest-Landau-level condition.

\par Ultimately, the theory may be generalised to the case where the $\nu^*_\alpha$'s can take any fractional value,
as to allow the multi-component generalisation of higher-generation FQHE states. An example of the latter in one
component is the $\nu=\tfracs{4}{11}$ state, which can be understood as a second generation FQHE state.
\cite{GoerbigEA2004EPL-GoerbigEA2004PRB-GoerbigEA2006,LopezFradkin2004-ChangJain2004}
\end{subsection}
\end{section}

\begin{section}{Singular transformations}
\label{sect_singular}
The analysis in the previous section demonstrates that a Chern--Simons transformation with a nonsingular charge matrix is already interesting in itself. However, transformations with singular charge matrices play an important role in the study of states with (partial) ferromagnetic order, since these states are described by singular exponent matrices.\cite{GoerbigRegnault2007} In this section, we investigate the consequences of the symmetry properties of the exponent matrices $M$ and $M^*$ and the charge matrix $K$ for the results of the previous section.

\begin{subsection}{Conditions on the ranks of the matrices}
Without performing the diagonalisation of the oscillator Hamiltonian, it is already possible to give some conditions on the exponent matrices and the charge matrix. Consider a state that is described by a singular exponent matrix $M$.
As a consequence, not all filling factors are defined separately. Suppose furthermore that the electronic and composite-fermion filling factors are given by $\sum_\beta M_{\alpha\beta}\nu_\beta=1$ and $\sum_\beta M^*_{\alpha\beta}\nu^*_\beta=1$, respectively, with $M=M^*+K$. We note that \eqn\eqref{eqn_lbstar} has to be satisfied simultaneously, which does not necessarily follow from the other conditions.\footnote{The given conditions are satisfied simultaneously if $M^*$ is a block-diagonal matrix where the entries within each block are equal to each other, which includes most physically interesting cases.}
From the fact that $M$, $M^*$ and $K$ are required to be nonnegative definite, it follows that also $K$ and $M^*$ are singular. More specifically, it follows that the null spaces of $M^*$ and $K$ may be of higher dimension than that of $M$.
As a consequence, the dimension of the null space of the exponent matrix is either increased or kept invariant by the Chern--Simons transformation. In other words, if before applying the Chern--Simons transformation the
theory involves a certain number of independent combinations of filling factors, then the number of independent
combinations after the transformation is either the same or lower. In terms of the ranks of the matrices, which is equal to their size minus the dimension of the null space (i.e., $\dim\ker M+\rank M=\kappa$), we find that the ranks of $K$ and $M^*$ must both be smaller than or equal to the rank of $M$.

\par For the case that $\rank M^* < \rank M$, which is not ruled out by the above discussion, some problems may arise. In this case, \eqn\eqref{eqn_lbstar} fixes the filling factors $\nu^*_\alpha$ to be confined to a subspace of the space of all solutions of $\sum_\beta M^*_{\alpha\beta}\nu_\beta^*=1$. For example, if $M=\bigl(\begin{smallmatrix}3&1\\1&3\end{smallmatrix}\bigr)$ and $K=\bigl(\begin{smallmatrix}2&0\\0&2\end{smallmatrix}\bigr)$, we have $M^*=\bigl(\begin{smallmatrix}1&1\\1&1\end{smallmatrix}\bigr)$, so that, based on the exponent matrices, the electronic and composite-fermion filling factors are given by $(\nu_1,\nu_2)=(1/4,1/4)$ and $\nu^*_1+\nu^*_2=1$, respectively. However, based on \eqn\eqref{eqn_lbstar}, the composite-fermion filling factors are fixed at $(\nu^*_1,\nu^*_2)=(1/2,1/2)$. Therefore, the matrix $M^*$ does not appropriately describe the possible composite-fermion filling factors of the system. We would expect that this leads to problematic results, if we used the Chern--Simons approach to obtain a separation between high-energy and low-energy degrees of freedom. For this reason, we will only analyse the case that $M$ and $M^*$ share their ranks. We stress that there is no problem in using a singular charge matrix $K$ if $M$ and $M^*$ are both nonsingular.
\end{subsection}

\begin{subsection}{The oscillator Hamiltonian}
\label{sect_singular_harmosc}%
Here, we discuss how the singularity of the matrix $K$ affects the analysis that we used to study the harmonic oscillator.
Apart from the zero modes in the harmonic oscillator, we must also take into account that
the number of independent constraints [\eqn\eqref{eqn_constraint_cp}] is reduced, since these also involve the
matrix $K$. Indeed, the number of independent constraints is given by the rank $r$ of the matrix $K$, whereas the
number of zero modes is $\kappa-r$.
Before we derive the fully general results, we find it instructive to illustrate the procedure first with a
simple example.

\par We consider a two-component system, where we choose the charge matrix of the Chern--Simons transformation
to be the singular matrix $K =\bigl(\begin{smallmatrix}2&2\\2&2\end{smallmatrix}\bigr)$.
The eigenvalues of $K$ are $4$ and $0$, and the respective eigenvectors are
$(1,1)/\sqrt{2}$ and $(1,-1)/\sqrt{2}$. We can write the constraints in components as
\begin{align*}
 0 &= \biggl(\ao_\alpha(\vq) - \frac{h}{e\abs{\vq}}\bigl(2\,\delta\rho_1(\vq)+2\,\delta\rho_2(\vq)\bigr)\biggr)\ket\phys,
\end{align*}
for $\alpha=1,2$. The two components $\ao_1$ and $\ao_2$ of the gauge field satisfy the same constraint, so that they are fixed to the fluctuations of the total density $\delta\rho_1+\delta\rho_2$.
On the other hand, the difference of density fluctuations $\delta\rho_1-\delta\rho_2$ (associated with the zero eigenvalue) is absent, implying that one may have zero-energy fluctuations that lower the particle number in one component while increasing that in the other component. Eventually such a reorganisation of the particles on the two components may even completely polarise the system, with $\nu_1=\nu$ and $\nu_2=0$. Inversely this means that in the case of a singular matrix $K$, we may always choose both filling factors nonzero or even equal, i.e., $N$ nonsingular, such as to render the harmonic approximation \eqref{eqn_hosc} valid.

\par We now turn to the harmonic oscillator Hamiltonian. We write $N=\diag(\nu_{1},\nu_{2})$, where $\nu_{{1,2}}$ are the electronic filling factors. For this example, we compute
\[
  E=\sqrt{N}\,K\sqrt{N}
  =2\,\begin{pmatrix}\nu_{1}&\sqrt{\nu_{1}\nu_{2}}\\ \sqrt{\nu_{1}\nu_{2}}&\nu_{2}\end{pmatrix}.
\]
This matrix is diagonalised as $C\Tp D C$, where $D=\diag(\lambda_1,\lambda_2)$ is the diagonal matrix with the eigenvalues $\lambda_1=2(\nu_1+\nu_2)$ and $\lambda_2=0$. The corresponding eigenvectors are proportional to $(\sqrt{\nu_1},\sqrt{\nu_2})$ and $(-\sqrt{\nu_2},\sqrt{\nu_1})$, respectively.
The diagonalised Hamiltonian is given by \eqn\eqref{eqn_hosc_diagonal__components}, where $\alpha=1,2$ and
\begin{align*}
   \begin{pmatrix}\abar_1\\ \abar_2\end{pmatrix}
   &= \frac{1}{\sqrt{\nu_{1}+\nu_{2}}}
     \begin{pmatrix}\nu_{1}\ao_1 + \nu_{2}\ao_2\\ \sqrt{\nu_{1}\nu_{2}}\,(-\ao_1+\ao_2)\end{pmatrix},\\
   \begin{pmatrix}\Pbar_1\\ \Pbar_2\end{pmatrix}
   &= \frac{1}{\sqrt{\nu_{1}+\nu_{2}}}
     \begin{pmatrix}\Po_1 + \Po_2\\ -\sqrt{\tfrac{\nu_{2}}{\nu_{1}}}\,\Po_1+\sqrt{\tfrac{\nu_{1}}{\nu_{2}}}\Po_2\end{pmatrix}.
\end{align*}
We note that $\Pbar_2$ is not present in the Hamiltonian since the term $\Pbar_2^\dagger\lambda_2^2\Pbar_2$ vanishes due to $\lambda_2=0$. The term $\abar_2^\dagger\abar_2$ also vanishes, since $\abar_2=0$, due to the constraint $\ao_1=\ao_2$. In the end, we obtain a
harmonic oscillator Hamiltonian with only one coordinate ($\abar_1$) and one momentum ($\Pbar_1$) component.

\par The Hamiltonian restricted to this single coordinate has a ground state
$\chi_{\mathrm{osc},1}=\exp[-\tfracsb{e}{2\hbar b}\abar^\dagger\psinv{D}\,\abar]$,
where we define $\psinv{D} = \diag(\tfrac{1}{2}(\nu_{1}+\nu_{2})^{-1},0)$. We note that $\chi_{\mathrm{osc},1}$ only involves $\abar_1$, but not $\abar_2$. Transforming back to the coordinates $(\ao_1,\ao_2)$ and imposing the constraints $\ao_1 = \ao_2 = \tfracsb{h}{e\abs{\vq}}\bigl(2\,\delta\rho_1+2\,\delta\rho_2\bigr)$, we obtain
\[
  \chi_{\mathrm{osc},1}
  =\exp\biggl(-\frac{1}{2}(\delta\rho_1+\delta\rho_2)^\dagger \frac{2\pi L^2}{\abs{\vq}^2}(2)\,(\delta\rho_1+\delta\rho_2)\biggr),
\]
where the notation $(2)$ is to point out that it should be interpreted as a matrix.
At this point, we observe that $(\delta\rho_1+\delta\rho_2)^\dagger(2)(\delta\rho_1+\delta\rho_2)$ is exactly equal to
$(\delta\rho_1,\delta\rho_2)^\dagger K(\delta\rho_1,\delta\rho_2)$. This means that in this example \eqn\eqref{eqn_chi_osc3_matrixform} is valid without change, and the other results concerning the Halperin wave functions hold as
well, as we have already mentioned in the discussion of the general oscillator function \eqref{eqn_chi_osc3_matrixform}.
We remark that the linear combination of filling factors $\nu_1-\nu_2$ is not present at all in the diagonalised theory.

\par Another important point is that we can make the connection with ferromagnetic Laughlin states in two-component systems.\cite{DeGailEA2008} For instance, the exponent matrix
$M=\bigl(\begin{smallmatrix}3&3\\3&3\end{smallmatrix}\bigr)$ defines a state for which the total filling factor is $\nu_1+\nu_2 = \tfracs{1}{3}$, but the separate filling factors are not defined, since the exponent matrix is singular.\cite{DeGailEA2008} Using the Chern--Simons transformation of the example above, we may understand this state in terms of a composite-fermion theory with exponent matrix $M^*=\bigl(\begin{smallmatrix}1&1\\1&1\end{smallmatrix}\bigr)$. This state
has total composite-fermion filling factor $\nu^*_1+\nu^*_2 = 1$, and again the separate filling factors are undefined. We remark that although the intermediate steps in the procedure contain the separate filling factors $\nu_1$ and $\nu_2$, the results are completely independent of $\nu_1-\nu_2$.

\par In contrast to the ferromagnetic Laughlin state discussed in the preceding paragraph, we may also discuss the two-component state at total filling factor $\nu=2/5$,
described by the matrix $M=\bigl(\begin{smallmatrix}3&2\\2&3\end{smallmatrix}\bigr)$ and the reduced exponent matrix $M^*=\bigl(\begin{smallmatrix}1&0\\0&1\end{smallmatrix}\bigr)$. Although the Chern--Simons transformation is described by a singular charge matrix $K$ and does therefore not impose a constraint on the relative particle distribution on the two components,
the constraint is imposed by $M^*$, $\nu_1^*=\nu_2^*=1$. The state thus described is then a spin-unpolarised state, as one could have also expected from the original exponent matrix $M$.

\par The reasoning given for the example above can be readily generalised to any situation in which $K$ is singular. Here, we assume that the rank $r$ of the matrix $K$ is smaller than the number of components $\kappa$. As argued in Appendix~\ref{sect_singular_gs}, the ground state can be decomposed as a product of
the usual ground state \eqref{eqn_chi_osc1_2_matrixform} restricted to the $r$ independent components, $\chi_{\mathrm{osc},r}$ and the degenerate part $\tilde\chi$ [see \eqns\eqref{eqn_ferro_(app)} and \eqref{eqn_chiosc_nondeg_(app)}]. Moreover, \eqn\eqref{eqn_chi_osc3_matrixform} remains valid even in the singular case, despite the fact that the original derivation involves the inverse of $K$. Hence, the Halperin connection in \sect\ref{sect_halperin_connection} is valid in the singular case without modification.

\par The equivalence of the decomposition \eqref{eqn_ferro_(app)} for the $\kappa$-component
oscillator wave function may be interpreted in a straightforward physical manner. Indeed, the decomposition
indicates that, in the case of a charge matrix $K$ of rank $r$, the ``reduced'' $r$-component wave function corresponds to an $r$-component Halperin wave function with gapped oscillator frequencies $\omega_{\alpha}$. The other factor $\tilde{\chi}$
in \eqn\eqref{eqn_ferro_(app)} corresponds to the $\kappa-r$ zero eigenvalues of the matrix $K$ with an associated space spanned by the oscillator components $\bar{a}_{\alpha}$, with $\alpha=r+1, \ldots, \kappa$. The ground-state manifold comprises therefore any possible combination of these components $\bar{a}_{\alpha}$, and a particular choice spontaneously breaks the residual
ground-state symmetry, which may be related to the ferromagnetic properties of the Halperin state, and $\tilde{\chi}$ may then be interpreted as the ferromagnetic part of the wave function.

\par In order to see this particular point, consider the $r$ constraints to fix the filling factors of the
first $r-1$ components. The last constraint then imposes simply the \emph{sum} of the fillings of all other components
$\alpha=r, \ldots, \kappa$. This is naturally a simplified assumption, because the $r$ constraints do not
in general fix particular components, but the dependencies may be more complicated.\footnote{ Other continuous subgroups of $\SU(\kappa)$ may appear as well as symmetry groups. These include product groups like $\SU(2)\times\SU(2)$, which is relevant for four-component systems with two $\SU(2)$ symmetries of a different origin, such as spin and pseudospin.} One is then free to distribute the involved
particles over these components in a quantum mechanical manner. All different distributions define the ground-state manifold.
Schematically, this may be formalised with the help of a wave function
\[
  \tilde{\chi}=u_{r}\ket{\alpha=r} + u_{r+1}\ket{\alpha = r+1} + \ldots + u_{\kappa}\ket{\alpha = \kappa},
\]
where the complex amplitudes $u_{\alpha}$ are subject to a normalisation condition, which plays the role of the last
constraint. These complex amplitudes may be viewed as the components of a $\mathrm{CP}^{\kappa-r}$ field.\footnote{The $\kappa-r+1$ complex components $u_{\alpha}$, $\alpha=r,\ldots,\kappa$ may indeed be viewed as an
element of the complex projective space $\mathrm{CP}^{\kappa-r}$ in which one identifies all elements that differ only by a
global (complex) factor $c$, $(u_r,\ldots,u_{\kappa}) \equiv (cu_r, \ldots, cu_{\kappa})$.} The ground-state manifold may then be described
by spatially constant $\mathrm{CP}^{\kappa-r}$ fields with a global
$\SU(\kappa - r +1)$ symmetry, which is precisely the symmetry group that describes the ferromagnetic properties of
the oscillator wave function.
In summary, this argument shows that, in the case of a Chern--Simons transformation with a matrix $K$
of rank $r$, one may decompose an arbitrary oscillator wave function into a product of a reduced $r$-component Halperin wave
function and a $\SU(\kappa - r+1)$-symmetric ferromagnetic part. Naturally, this symmetry may be further reduced if the
components of the Chern--Simons field fix further filling factors.

\par We finally mention that the spontaneous breaking of the $\SU(\kappa - r+1)$ symmetry yields Goldstone modes, which are
physical (pseudo)spin waves. On the level of the Gaussian model, these Goldstone modes are dispersionless and remain at zero
energy. This is no longer the case if one takes into account interactions between the particles associated with the different components.
One may indeed treat rather easily a density-density interaction within the present model. This interaction may be translated, via the
constraints \eqref{eqn_constraint_cp}, into an interaction between the oscillator fields, which one can then diagonalise within the
Gaussian model. Notice that these fields are coupled to the
$\vec{\Pi}_{\alpha}$ [see \eqn\eqref{eqn_hcoupl}], which describe the low-energy electronic degrees of freedom. A discussion of collective Goldstone-type
modes is therefore more involved and requires a decoupling of the oscillator and the electronic degrees of freedom. However,
the Chern--Simons analysis within the Gaussian model yields valuable insight into the ferromagnetic properties of the states, which
are governed by symmetry, as well as into the number of their Goldstone modes.
\end{subsection}
\end{section}

\begin{section}{Conclusions}
\label{sect_conclusions}
In conclusion, we have studied a microscopic Chern--Simons approach to general multi-component quantum Hall systems (with $\kappa$
components). Beyond the mean-field approximation, which yields a renormalisation of the magnetic field that depends
on the average particle densities for each component, their fluctuations are taken into account within a Gaussian model
of auxiliary gauge fields. These gauge fields, introduced by Shankar and Murthy in the framework of the Hamiltonian theory of the
FQHE,\cite{ShankarMurthy1997,MurthyShankar1998,Shankar1999,Shankar2001,MurthyShankar2003} are indeed connected via constraints
to the component density fluctuations.

\par The analysis of the Gaussian model ---although it may be viewed as a first step in the discussion of a more complete Hamiltonian
theory for multi-component quantum Hall systems--- already yields valuable insight into the structure and the correctness of the
Chern--Simons theory, which is characterised by a symmetric $\kappa\times\kappa$ charge matrix
$K$.\cite{WenZee1992PRB46-WenZee1992PRL69} Most saliently, one needs to discard charge matrices with negative
eigenvalues because the associated Chern--Simons theories yield oscillator ground-state wave functions that are not normalised.
This is in line with physical insight obtained from a multi-component version of Laughlin's plasma
picture\cite{Laughlin1983} according to which charge matrices with negative eigenvalues yield inhomogeneous ground states
where the components phase-separate.\cite{DeGailEA2008}

\par Whereas singular charge matrices, with zero eigenvalues, had originally been discussed by Lopez and Fradkin\cite{LopezFradkin1995} only for the $\SU(2)$-symmetric case,
we have argued here that the associated Chern--Simons theories reflect underlying ferromagnetic states in a more general setting. Indeed, we have
shown that the density fluctuations of the $\kappa$ components are then determined by only $r<\kappa$ constraints, such that
$\kappa-r$ particular combinations of the component densities may be chosen freely in the ground-state manifold, which
is thus described by the $\SU(\kappa-r+1)$ group. This symmetry is spontaneously broken by a particular ferromagnetic
state, which can be described by $\kappa-r$ different Goldstone modes that may be viewed as generalised spin waves. Our results
encompass the particular $\SU(2)$ case of two-component Chern--Simons theories discussed in the
literature.\cite{WenZee1991PRB44,WenZee1992PRB46-WenZee1992PRL69,LopezFradkin1995,Rajaraman1997,ScarolaJain2001}

\par We emphasise moreover that the analysis of the microscopic multi-component Chern--Simons theory within the Gaussian approximation
heuristically yields trial wave functions for multi-component quantum Hall systems that may be further studied numerically. As
an example, we have discussed generalised $\kappa$-component Halperin wave functions that play a similarly central role
as Laughlin's wave functions do in one-component quantum Hall systems. Beyond these generalised Halperin wave functions, we have
briefly discussed a second class of states, where the residual wave function that is not encoded in the Chern--Simons oscillator
part $\chi_\mathrm{osc}$ is a product of Slater determinants of $p_{\alpha}$ completely filled ($\alpha$-component) composite-fermion
levels. This construction is reminiscent of Jain's generalisation of one-component Laughlin wave functions to filling factors
$\nu=p/(2sp+1)$.\cite{Jain1989,Jain1990,LopezFradkin1991}

\end{section}
{}
\begin{acknowledgments}
We are grateful to G. Murthy and N. Regnault for useful discussions.
This work was partially supported by the Netherlands Organisation for Scientific Research (NWO). We further acknowledge support
from the ANR under Grant No. ANR-JCJC-0003-01.
\end{acknowledgments}
{}
\appendix

\begin{section}{Multi-component plasma analogy}
\label{sect_plasma_analogy_derivation}%
The single-component plasma analogy proposed by Laughlin\cite{Laughlin1983} is readily generalised to the multi-component case. Here, we use the ground state
\begin{equation}\label{eqn_chi_osc3}
  \chi_\mathrm{osc}
   = \exp\biggl(-\frac{1}{2}\sum_\vq\sum_{\beta,\gamma}\delta\rho_\beta(-\vq) \frac{2\pi L^2}{\abs{\vq}^2}K_{\beta\gamma}\delta\rho_\gamma(\vq)\biggr),
\end{equation}
which is \eqn\eqref{eqn_chi_osc3_matrixform} written out in components. Recalling that $2\pi L^2/\abs{\vq}^2$ is the Fourier transform of
$-\!\log\abs{\vr}$, we perform an inverse Fourier transformation and we substitute the density fluctuations $\delta\rho_\alpha(\vr)=\sum_\ja\delta(\vr-\vr_\ja)-\nel{\alpha}$. Then, we can rewrite $\chi_\mathrm{osc}$ as
\begin{widetext}
\[
  \chi_\mathrm{osc}
   = \exp\Biggl[\frac{1}{2}
     \sum_{\alpha,\beta}K_{\alpha\beta}\int\intd^2\vr\intd^2\vr'
       \biggl(\sum_{j_\alpha=1}^{N_\alpha}\delta(\vr-\vr_{j_\alpha})-\nel{\alpha}\biggr)
       \log\abs{\vr-\vr'}
       \biggl(\sum_{k_\beta=1}^{N_\beta}\delta(\vr'-\vr_{k_\beta})-\nel{\beta}\biggr)
   \Biggr].
\]
By evaluating the integrals, one finds
\begin{equation*}
  \chi_\mathrm{osc}
  =\mathrm{const}\cdot
  \prod_{\alpha,\beta}\prod_{\substack{j_\alpha,k_\beta\\j_\alpha\neq k_\beta}}\abs{\vr_{j_\alpha}-\vr_{k_\beta}}^{K_{\alpha\beta}/2}
  \exp\biggl(-\frac{\pi}{2}\sum_{\alpha,\beta}\nel{\alpha}K_{\alpha\beta}\sum_{k_\beta}\abs{\vr_{k_\beta}}^2\biggr).
\end{equation*}
Using that $\nu_\alpha = 2\pi\lB^2\nel{\alpha}$, and changing to complex notation, with $z=x-iy$,\footnote{This counterintuitive definition is used in order to have analytic lowest Landau level wave functions. This is due to the negative charge $-e$ of the electrons, for which the basic Hamiltonians are defined. Therefore, one has $\theta(\vr)=\arg(x+\ii y)=\arg z^*= -\arg(z)$.} we can explicitly write this expression as
\begin{equation*}
  \chi_\mathrm{osc}
  =\mathrm{const}\cdot
  \prod_{\alpha}\prod_{\substack{j_\alpha,k_\alpha\\j_\alpha < k_\alpha}}\abs{z_{j_\alpha}-z_{k_\alpha}}^{K_{\alpha\alpha}}
  \prod_{\substack{\alpha,\beta\\ \alpha<\beta}}\prod_{j_\alpha, k_\beta}\abs{z_{j_\alpha}-z_{k_\beta}}^{K_{\alpha\beta}}
  \exp\biggl(-\sum_{\alpha,\beta}\nu_{\alpha}K_{\alpha\beta}\sum_{k_\beta}\frac{\abs{z_{k_\beta}}^2}{4\lB^2}\biggr).
\end{equation*}
The Jastrow-like products in this expression only contain distances between
the particles, i.e., only the moduli $\abs{z_{j_\alpha}-z_{k_\beta}}$. Phase factors of the form  $[(z_{j_\alpha}-z_{k_\beta})/\abs{z_{j_\alpha}-z_{k_\beta}}]^{K_{\alpha\beta}}=\exp[\ii K_{\alpha\beta} \arg(z_{j_\alpha}-z_{k_\beta})]=\exp[-\ii K_{\alpha\beta}\theta(\vr_{j_\alpha}-\vr_{k_\beta})]$ are obtained from substitution of the full density $\rho_\alpha(\vr)=\sum_\ja\delta(\vr-\vr_\ja)$ into the Chern--Simons transformation \eqref{eqn_cs_definition}.
Applying this transformation to $\chi_\mathrm{osc}$, we obtain the product of the latter with the phase factors,
\begin{equation}\label{eqn_wf_halperin1_(app)}
    \psi(\{z_\ja\})=
    \prod_{\alpha}\prod_{\substack{j_\alpha,k_\alpha\\j_\alpha < k_\alpha}}(z_{j_\alpha}-z_{k_\alpha})^{K_{\alpha\alpha}}
    \prod_{\substack{\alpha,\beta\\ \alpha<\beta}}\prod_{j_\alpha, k_\beta}(z_{j_\alpha}-z_{k_\beta})^{K_{\alpha\beta}}
    \exp\biggl(-\sum_{\alpha,\beta}\nu_{\alpha}K_{\alpha\beta}\sum_{k_\beta}\frac{\abs{z_{k_\beta}}^2}{4\lB^2}\biggr)
    \,\phi_{\{\nu_\alpha^*\}}(\{z_\ja\}),
\end{equation}
where $\phi_{\{\nu_\alpha^*\}}$ denotes the composite-particle
wave function for filling factors $\nu^*_\alpha$, which will be investigated in the following. The magnetic lengths
appearing in $\phi_{\{\nu_\alpha^*\}}(\{z_\ja\})$ are the reduced magnetic lengths $\lBB{B^*_\alpha}$ given by \eqn\eqref{eqn_lbstar}.

\par As an example, we consider the situation in which $\nu^*_\alpha$ can be determined by an exponent matrix $M^*$,\cite{WenZee1992PRB46-WenZee1992PRL69,GoerbigRegnault2007,DeGailEA2008} such that $\phi_{\{\nu_\alpha^*\}}$ is the Halperin wave function
\begin{equation}\label{eqn_phi_nustar_halperin_(app)}
  \phi_{\{\nu_\alpha^*\}}(\{z_\ja\})
  =\prod_{\alpha}\prod_{\substack{j_\alpha,k_\alpha\\j_\alpha < k_\alpha}}(z_{j_\alpha}-z_{k_\alpha})^{M^*_{\alpha\alpha}}
  \prod_{\substack{\alpha,\beta\\ \alpha<\beta}}\prod_{j_\alpha, k_\beta}(z_{j_\alpha}-z_{k_\beta})^{M^*_{\alpha\beta}}
  \exp\biggl(-\sum_\alpha\sum_{k_\alpha}\frac{\abs{z_{k_\alpha}}^2}{4\lBB{B^*_\alpha}^2}\biggr).
\end{equation}
Combining \eqns\eqref{eqn_wf_halperin1_(app)} and \eqref{eqn_phi_nustar_halperin_(app)}, we obtain the full electronic wave function,
\begin{equation}\label{eqn_wf_halperin2_(app)}
  \psi(\{z_\ja\})
  =\prod_{\alpha}\prod_{\substack{j_\alpha,k_\alpha\\j_\alpha < k_\alpha}}(z_{j_\alpha}-z_{k_\alpha})^{K_{\alpha\alpha}+M^*_{\alpha\alpha}}
   \prod_{\substack{\alpha,\beta\\ \alpha<\beta}}\prod_{j_\alpha, k_\beta}(z_{j_\alpha}-z_{k_\beta})^{K_{\alpha\beta}+M^*_{\alpha\beta}}
   \exp\biggl(-\sum_\alpha\sum_{k_\alpha}\frac{\abs{z_{k_\alpha}}^2}{4\lB^2}\biggr),
\end{equation}
\end{widetext}
which is the Halperin wave function for the exponent matrix $M_{\alpha\beta}=M^*_{\alpha\beta}+K_{\alpha\beta}$.\cite{GoerbigRegnault2007}
Here, we have expressed the effective magnetic lengths in the exponential of \eqn\eqref{eqn_phi_nustar_halperin_(app)} in terms of the original one, as $1/\lBB{B^*_\alpha}^2 = (1-\sum_\beta K_{\alpha\beta}\nu_\beta)/\lB^2$, by virtue of \eqn\eqref{eqn_lbstar}.
\end{section}

\begin{section}{Ground state in the singular case}
\label{sect_singular_gs}
The reasoning given for the two-component example in \sect\ref{sect_singular_harmosc} can be extended to any number of components. Suppose that the charge matrix $K$ (being a
$\kappa\times \kappa$ symmetric nonnegative definite matrix) is of rank $r$, which means that it has $r$ independent
rows or columns. In particular, there are $\kappa-r$ rows or columns that can be written as a linear combination of the
other $r$ independent rows or columns. This also means that the dimension of the null space, or equivalently, the
multiplicity of zero eigenvalues is equal to $\kappa-r$.

\par Since the constraints \eqref{eqn_constraint_cp} are expressed as a linear relation involving the matrix $K$, there are only $r$ independent
constraints. Hence, the vector $\ao=(\ao_1,\ldots,\ao_\kappa)$ lives only in an $r$-dimensional subspace; $\kappa-r$ of
its components can be written as a linear combination of the other $r$.

\par Now we analyse the Hamiltonian \eqref{eqn_hosc}. Since we have assumed that the matrix of densities $N$ is nonsingular
(i.e., all filling factors are nonzero, as required for the harmonic approximation to be valid), the rank of $E=\sqrt{N}\,K\sqrt{N}$ is equal to the rank of $K$. This means that $E$ has $r$ positive
eigenvalues and $\kappa-r$ zero eigenvalues, just as the matrix $K$. We diagonalise $E$ as usual in terms of a diagonal
matrix $D$ and an orthogonal matrix $C$ such that $E=C\Tp D\,C$. Note that the order of the eigenvalues on the diagonal
of $D$ (and simultaneously the order of the rows of $C$) may be chosen at will, so that we may choose for simplicity
$D=\diag(\lambda_1,\ldots,\lambda_r,0,\ldots,0)$, where $\lambda_1,\ldots,\lambda_r$ are the positive eigenvalues of
$E$. In the diagonalised Hamiltonian \eqref{eqn_hosc_diagonal_}, the components $\Pbar_{r+1},\ldots,\Pbar_\kappa$ are
absent since they are multiplied with the zero eigenvalues of $D$. We still have $\kappa$ components of $\abar$ in the
Hamiltonian, but we should remember that only $r$ of them are independent.

\par The diagonalised Hamiltonian contains $r$ nonzero eigenvalues, which depend on the filling factors $\nu_\alpha$.
However, some variations of the filling factors will leave the eigenvalues, and hence the diagonalised Hamiltonian,
invariant, namely those satisfying the equation
\[
  0 = (\nabla\lambda_\beta)\cdot\delta\nu = \sum_\alpha \ddau{\lambda_\beta}{\nu_\alpha}\delta\nu_\alpha
  \quad\text{for all $\beta$}.
\]
In other words, the desired variations are the vectors in the null space of the gradient matrix $(\nabla\lambda)$ of
the eigenvalues, which is defined as the matrix of derivatives of $\lambda$, with respect to $\nu$,
$(\nabla\lambda)_{\beta\alpha}=\tfddau{\lambda_\beta}{\nu_\alpha}$. Since $\kappa-r$ of the eigenvalues $\lambda_\beta$
are zero, the rank of the gradient matrix is at most $r$, and this consequently means that we can find at least
$\kappa-r$ independent variations of the filling factors which leave the eigenvalues invariant.

\par In the example discussed in \sect\ref{sect_singular_harmosc},
we observed that $\nu_1-\nu_2$ does not appear in the diagonalised Hamiltonian.
In order to demonstrate the procedure sketched in the preceding paragraph, we compute the gradients of the eigenvalues
$0$ and $2(\nu_{1}+\nu_{2})$. Obviously, in this example the gradient matrix is
$(\nabla\lambda)=\bigl(\begin{smallmatrix}0&0\\2&2\end{smallmatrix}\bigr)$
and its null space is spanned by the single vector $(1,-1)$. Since this vector is independent of the filling
factors $\nu_{1}$ and $\nu_{2}$, we can state that all eigenvalues, and hence the diagonalised Hamiltonian, are invariant under the
transformation $\{\nu_1\to\nu_1+\delta\nu,\nu_2\to\nu_2-\delta\nu\}$. This means that the linear combination $\nu_1-\nu_2$ is
completely absent from the Hamiltonian, as argued earlier. We remark that the variation that leaves the Hamiltonian invariant need not always be
constant in the filling factors $\nu_\alpha$. However, for physically relevant systems, the variations are constant, describing particle exchange among different components.

\par We now return to the diagonalised Hamiltonian, and try to find the lowest-energy states, in the same way as we have done for the example in \sect\ref{sect_singular_harmosc}. For the moment,
we do not impose the constraints, thus regarding all components $\abar_\alpha$ as independent. Only the first $r$
components
$\abar_{\tilde\alpha}$ ($\tilde\alpha=1,\ldots,r$) have a corresponding momentum operator
$\Pbar_{\tilde\alpha}$ in the Hamiltonian, while the other $\kappa-r$ components do not. This means that the
resulting states are degenerate in the coordinates $\abar_{r+1},\ldots,\abar_{\kappa}$. Thus, we may write the
lowest-energy states as
\begin{equation}\label{eqn_ferro_(app)}
  \chi_\mathrm{osc}(\abar_1,\ldots,\abar_\kappa)
  =\chi_{\mathrm{osc},r}(\abar_1,\ldots,\abar_r)\,\tilde\chi(\abar_{r+1},\ldots,\abar_{\kappa}),
\end{equation}
where $\tilde\chi$ is the degenerate part of the wave function (further discussed in \sect\ref{sect_singular_harmosc}), and
\begin{equation}\label{eqn_chiosc_nondeg_(app)}
\chi_{\mathrm{osc},r}(\abar_1,\ldots,\abar_r)
  = \exp\biggl(
     -\frac{e}{2\hbar b} \sum_{\tilde\alpha=1}^r\abar_{\tilde\alpha}^\dagger\lambda_{\tilde\alpha}^{-1}\abar_{\tilde\alpha}
   \biggr)
\end{equation}
is the nondegenerate part. Notice that the components associated with the zero eigenvalues of the matrix $E$ do not contribute. By the
observation that the pseudoinverse\cite{Penrose1955-BenIsraelGreville1974} of
$D=\diag(\lambda_1,\ldots,\lambda_r,0,\ldots,0)$ is equal to
$\psinv{D}=\diag(\lambda_1^{-1},\ldots,\lambda_r^{-1},0,\ldots,0)$, we may also rewrite $\chi_{\mathrm{osc},r}$ as
\begin{equation*}
  \chi_{\mathrm{osc},r}
   = \exp\biggl(-\frac{e}{2\hbar b}\abar^\dagger\psinv{D}\,\abar\biggr)
   =\exp\biggl(-\frac{e}{2\hbar b}\ao{}^\dagger \psinv{K}\ao\biggr),
\end{equation*}
where we used that $\abar^\dagger \psinv{D}\, \abar=\ao{}^\dagger \psinv{K}\ao$. This result is nothing else than \eqn\eqref{eqn_chi_osc1_2_matrixform} with the inverses of $D$ and $K$ replaced by the pseudoinverses. Substituting the
density fluctuations $\delta\rho_\alpha$ for the gauge fields $\ao_\alpha$ using the constraint
\eqref{eqn_constraint_cp} yields exactly \eqref{eqn_chi_osc3_matrixform}
by virtue of the property of $\psinv{K}$ that $K\,\psinv{K}\,K = K$. This result is exactly equal to the steps we followed before, but only with $K^{-1}$ replaced by $\psinv{K}$. Therefore, the ground state \eqref{eqn_chi_osc3_matrixform} found for
the case of strictly positive eigenvalues is also valid if there are zero eigenvalues. All subsequent steps concerning the connection to the trial wave functions remain valid as well.
\end{section}


\end{document}

%% file: _macros-math-revtex.tex
\newcommand{\A}          {\mathcal{A}}

\newcommand{\diag}       {\mathop{\mathrm{diag}}       \nolimits}

\newcommand{\rank}       {\mathop{\mathrm{rank}}       \nolimits}

\newcommand{\SU}         {\mathop{\mathbf{SU}}         \nolimits} 

\newlength{\extarr }  
\newlength{\prtarr }  
\newlength{\dprtarr}  
\newlength{\limarr }  
\newlength{\sumarr }  
\newlength{\intarr }  
\newlength{\matarr }  
\newlength{\dmatarr}  
\newlength{\ddetarr}  

\newlength{\extquo }
\newlength{\prtquo }
\newlength{\dprtquo}
\newlength{\limquo }
\newlength{\sumquo }
\newlength{\intquo }
\newlength{\matquo }
\newlength{\dmatquo}
\newlength{\ddetquo}



%% file: _macros-common.tex
\newlength{\muspace}
\newcommand{\comm}[2]                   {[#1,#2]}                            

\newcommand{\ddau}[2]                   {\frac{\partial #1}{\partial #2}}    
\newcommand{\ee}                        {e}                                  
\newcommand{\ii}                        {i}                                  
\newcommand{\intd}                      {\mathrm{d}}                         
\newcommand{\inv}                       {^{-1}}                              

\newcommand{\Tp}                        {^\mathrm{T}}                        

\newcommand{\what}[1]                   {\widehat{#1}}                       
\newcommand{\wbar}[1]                   {\overline{#1}}                      

\renewcommand{\vec}[1]                  {\mathbf{#1}}                        

\newcommand{\unitvec}[1]                {\vec{e}_{#1}}                       



%% file: _macros-phys.tex
\newcommand{\avg}[1]                    {\langle#1\rangle}                   
\newcommand{\psinv}[1]                  {\what{#1}}                          

\newcommand{\lB}                        {l_B}                                
\newcommand{\lBB}[1]                    {l_{#1}}                             
\newcommand{\omegac}                    {\omega_\mathrm{c}}                  
\newcommand{\CS}                        {\mathrm{CS}}                        
\newcommand{\CP}                        {\mathrm{CP}}                        
\newcommand{\nel}[1]                    {n_{#1}}                             

\newcommand{\tfracs}[2]                 {\tfrac{#1}{#2}}                     
\newcommand{\tfracsb}[2]                {\tfrac{#1}{#2}}                     
\newcommand{\tfraci}[2][1]              {\tfrac{#1}{#2}}                     
\newcommand{\tfddau}[2]                 {\ddau{#1}{#2}}                      

